\documentclass{article}

\usepackage{arxiv}

\usepackage[utf8]{inputenc} 
\usepackage[T1]{fontenc}    
\usepackage{hyperref}       
\usepackage{url}            
\usepackage{booktabs}       
\usepackage{amsmath}
\usepackage{amsfonts}       
\usepackage{nicefrac}       
\usepackage{microtype}      
\usepackage{lipsum}
\usepackage{graphicx}
\graphicspath{ {./images/} }

\title{Phase tomography with axial structured illumination}

\author{
  Nishant Goyal \\
  Department of Physics,\\ Indian Institute of Technology Delhi,\\ New Delhi 110016 India \\
  \texttt{phz218060@iitd.ac.in} \\
   \And
  Kedar Khare \\
  Optics and Photonics Centre,\\
  Institute of Technology Delhi,\\ New Delhi 110016 India\\
  \texttt{kedark@iitd.ac.in} \\
}

\begin{document}
\maketitle
\begin{abstract}
Holographic Tomography (HT) or Optical Diffraction Tomography (ODT) provides slice-by-slice information about the refractive index (RI) of three-dimensional (3D) samples and is emerging as an important label-free imaging modality for Life sciences. HT systems go beyond the digital holographic microscopy (DHM) systems that provide a two-dimensional (2D) representation of the total accumulated phase acquired by a plane beam on transmission through a 3D sample. While the early HT systems used a direct reconstruction methodology based on the Fourier diffraction theorem, in recent years, there has been an increasing shift towards using iterative optimization frameworks for solving the 3D RI reconstruction problem. Iterative frameworks naturally offer several advantages for addressing the data incompleteness issues (e.g., missing illumination angles) and have superior noise handling capability, since they employ suitable constraint functions. Despite this algorithmic framework shift, the HT system hardware still largely uses the multi-angle illumination geometries that were suitable for reconstructions based on the Fourier diffraction theorem. The present work examines the possibility of HT reconstruction through the use of on-axis structured illumination(s) that nominally illuminates the 3D sample just along the axial direction. Through a simulation study, it is shown that a cross-talk free slice-by-slice 3D RI reconstruction of the sample is possible in this case via the use of sparsity penalties if the slice-to-slice distance obeys a design curve based on the notion of effective depth of focus. The simulation results for two-, three- and four-slice 3D objects with laterally overlapping features clearly outline the separate roles played by the slice-to-slice de-correlation of the field propagating through the 3D sample (modeled via multi-slice beam propagation) and that of the sparsity penalty used to guide the iterative solution. Our results suggest the possibility of realizing an Axial Structured Illumination Tomography (ASIT) system configuration that avoids the use of hardware-intensive multi-angle illumination geometry. An ASIT system can, for example, be achieved by minimal modification of a traditional digital holographic microscope (DHM) system.
\end{abstract}


\section{Introduction}
Tomographic imaging of transparent objects has become an active area of research and is variously referred to as optical diffraction tomography (ODT) or holographic tomography (HT). HT systems aim to provide 3-dimensional (3D) refractive index distribution of transparent objects like cells. The phase map generated by traditional digital holographic microscopy (DHM) systems is approximately a 2-dimensional (2D) projection of the refractive index. While the reconstructed phase function in DHM imaging is often rendered as a surface relief map, it does not have tomographic information. HT systems can potentially provide more detailed information about the internal structure of transparent or unstained 3D objects. They have generated much interest among bio-science research community as there are almost no parallel label-free optical technologies for obtaining tomographic information. Following Wolf's early treatment \cite{wolf1969three} of the ODT problem, the HT systems have largely used a system approach where a number of digital holograms (often 100s in number) of the 3D object are recorded by varying the direction of illumination. The 2D phase maps recovered from these digital holograms with illumination angle diversity are then used to reconstruct a 3D refractive index (RI) map \cite{lin2017optically, tian20153d}. The reconstruction methodology has parallels with the developments in the area of X-ray Computed Tomography, with the key difference being that it is now based on the Fourier diffraction theorem rather than the Fourier slice theorem. This multi-angle illumination methodology is now maturing into commercial technology for generic use by bio-science researchers. Tilt-series (similar to multi-angled illuminations) has also been widely used \cite{chreifi2019rapid,fernandez2018cryo} in other regimes like electron cryotomography. 

The aim of the present work is to examine the HT problem and understand if multiple angled illuminations can be replaced by axial structured illumination(s) to achieve good quality tomographic reconstructions with minimal slice-to-slice cross talk. By the term 'axial,' we refer to an illumination wavefront whose nominal propagation direction is along the optic axis (or z-axis) of the imaging system. Axial illumination(s) with phase structuring can be practically implemented in a manner that avoids the hardware (and calibration) complexity associated with recording of large number of digital holograms at various illumination angles. From a system building perspective, ASIT can be implemented with marginal hardware changes to existing DHM systems. Our work is inspired by prior investigations \cite{birdi2020true,rajora20233d} that used single-scattering approximation and were aimed at understanding the nature and limits of 3D image formation in single-view digital holography when axial plane wave illumination is used. In these works, it was observed that laterally separated phase objects at different depths could be reconstructed well, however, the reconstruction of laterally overlapping objects was only approximate even when the objects were gradient-sparse. The main theme of our investigation is to examine if a smaller number of axial structured illuminations can provide the same diversity of information as provided by multi-angled illuminations. If the information diversity in the two cases is similar, then an optimization-based reconstruction utilizing image sparsity ideas should be able to reconstruct tomographic images in either case. Our results suggest that for a given axial tomographic resolution, achieving a cross-talk free tomographic reconstruction depends on two considerations: (i) the spatial bandwidth of the exit wave from a typical layer of the 3D object, and (ii) the dimensionality mismatch between number of unknowns and number of measurements as allowed by techniques such as compressive imaging. We present a simulation study aimed to provide design guidelines for assessing the feasibility of tomographic reconstruction with axial structured illumination(s).

The paper is organized as follows. In Section \ref{s2}, we briefly summarize the traditional HT technique based on the Fourier diffraction theorem. In Section \ref{s3}, we discuss the axial structured illumination configuration of ASIT system where we further describe the reconstruction methodology, the special case of single plane wave illumination, and the guidelines for selecting the structured illumination of appropriate spatial bandwidth and the associated de-correlation length of the propagating fields as a design guideline to specify the axial slice separation. The simulation study is designed to clearly illustrate the role of the de-correlation distance and the sparsity penalty. Finally, in Section \ref{s4}, we provide discussion and limitations of the present study, future directions, and concluding remarks.

\section{Holographic tomography with multi-angled illuminations}\label{s2}
\begin{figure}[tbp]
    \centering
    \includegraphics[width=0.8\textwidth]{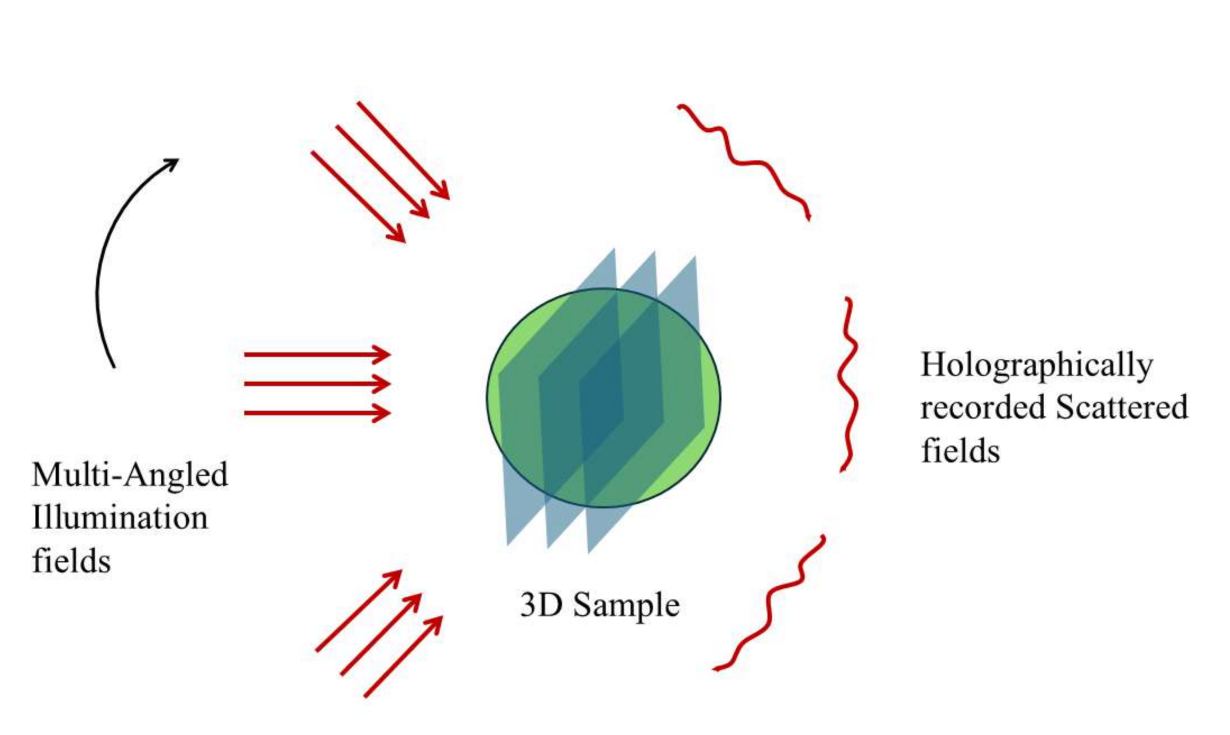}
    \caption{Multi-angled illuminations based holographic tomography}
    \label{fig:Multi-angled}
\end{figure}

While the subject of holographic tomography has been widely studied in the literature, we provide a brief summary of the main results for completeness. Holographic tomography (HT) or Optical Diffraction Tomography (ODT), originally proposed by Wolf \cite{wolf1969three},  aims to reconstruct the three-dimensional refractive index distribution of test samples by capturing and analyzing coherently scattered light from the object. Approximations such as the Born model have historically been employed to simplify the complex problem of HT by assuming that the scattered field is weak enough that multiple scattering effects can be neglected. Under the Born approximation, the weakly scattered field $u_{scat}(\vec{r})$ can be given as \cite{sung2009optical}:
\begin{equation}\label{eq:ODT1}
    u_{scat} (\vec{r}) = \int d^3\vec{r'}\ O(\vec{r'})\ u_{inc}(\vec{r'})\ G(\vec{r} - \vec{r'}), 
\end{equation}
where $O(\vec{r}) = k^2 [n^2_{s}(\vec{r}) - n^2_{m}]$ is object function and $G(|\vec{r}|) = \exp(i n_{m} k |\vec{r}|) / (4 \pi |\vec{r}|)$ is the associated Green's function for propagation. The quantities $n_{s}(\vec{r})$ and $n_{m}$ are the refractive index (RI) values of the sample, and the surrounding medium, respectively, and $k = 2\pi/\lambda$ is the free-space wave number for the wavelength of illumination $\lambda$. It is possible to establish a Fourier transform relation based on (\ref{eq:ODT1}), which is commonly referred to as the Fourier diffraction theorem. If the scattered field $u_{scat}(\vec{r})$ is measured on the plane at $z = z_P$ beyond the 3D sample, we have
\begin{equation}\label{eq:ODT2}
    \tilde{O} (k_x, k_y, k_z) = \frac{i k_z}{\pi} \tilde{u}_{scat} (k_x, k_y; z = z_P). 
\end{equation} 
Here $\tilde{O}(k_x, k_y, k_z)$ and $\tilde{u}_{scat}$ are the 3D and 2D Fourier transforms of $O$ and $u_{scat}$ respectively and $k_x^2 + k_{y}^2 + k_{z}^2 = k^2$. This relation is analogous to the Fourier projection theorem used in X-ray computed tomography. The Fourier diffraction theorem above essentially states that for every angled plane wave illumination, $\tilde{u}_{scat}$ can be mapped onto a hemispherical surface in the Fourier domain within limits set by the numerical aperture (NA) of the experimental system. By collecting the scattered field data from multiple angles (shown in Figure \ref{fig:Multi-angled}) and filling the 3D Fourier space, it is possible to obtain the volumetric RI distribution of the sample using a 3D inverse Fourier transformation of $\tilde{O}$. Detailed mathematical derivations for this result can be found in prior literature \cite{sung2009optical, balasubramani2021holographic,kim2014high}. While the theorem requires illumination from all angles in 3D space, the experimental data is often incomplete in this respect leading to the missing cone artifacts \cite{lim2015comparative} in the reconstructed image.
A multi-angle illumination-based HT configuration is conceived based on the basic results described above. The problem of number of illuminations required has also been studied \cite{fannjiang2022uniqueness} where it was proved that $n + 1$ diffraction measurements are sufficient for a $n \times n \times n$ object grid under Born approximation. The above framework does not, however, include image sparsity ideas that are integral to currently used image reconstruction methods. While HT reconstruction literature now increasingly makes use of such optimization-based reconstruction methods \cite{chowdhury2019high,chen2020multi,kamilov2016optical, kus2015active}, the basic geometry of HT systems has still remained the same. In \cite{birdi2020true} it was shown that plane wave illumination is able to perform tomographic reconstruction for laterally separated discrete 3D objects at different depths by use of a TV based reconstruction. When a similar framework was applied to a label-free 3D object (red blood cell) with laterally overlapping features, the reconstruction with single axial plane wave illumination was found to be approximate \cite{rajora20233d}. An additional focusing criterion was used in this work as a z-dependent weight in the iterations to help restrict the axial extent of the test object. It is, however, clear that in a single-view plane wave illumination configuration, it is difficult to provide tomographic reconstruction of laterally overlapping objects even if sparsity constraints are employed in the reconstruction. Our aim here is to examine if single-view recording configuration with axial structured illumination(s) can provide a good quality tomographic reconstruction with minimal slice-to-slice cross-talk. As we will explain in the next section, the key idea behind this configuration is that, unlike plane wave illumination, structured illumination evolves while propagating through the depth of the object. Under single-scattering approximation, this implies that individual slices of the object are illuminated by distinct wavefronts. The recorded holographic data beyond the 3D object, therefore, has much more diversity, which we hope to exploit in the sparsity-based iterative reconstruction process. In the more accurate beam propagation model, i.e., MSBP, that we use for reconstructions in the later sections of this work, the propagating field is affected by the material in each slice of the object, thus further increasing this diversity. Our aim is to show the importance of considering both the field de-correlation between adjacent planes of the 3D object and the use of sparsity penalty in realizing a cross-talk free 3D RI reconstruction.

\section{Problem configuration of ASIT}\label{s3}
\begin{figure}[tbp]
    \centering
    \includegraphics[width=0.8\textwidth]{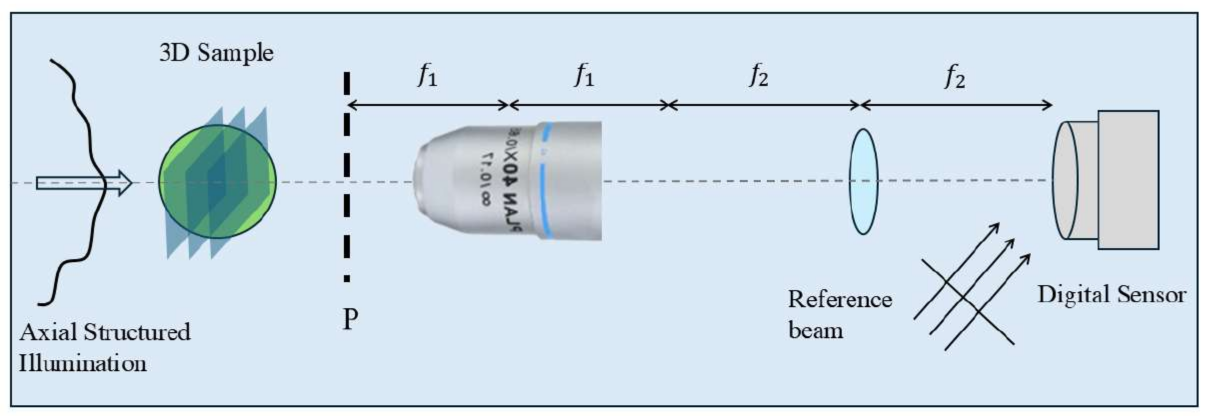}
    \caption{Schematic Setup of ASIT system. Axial structured field illuminates the sample, and the scattered field is holographically recorded (using a tilted reference beam) at plane P using a 4f Imaging system. $f_1$ and $f_2$ are not drawn to scale. }
    \label{fig:schematic}
\end{figure}

The ASIT configuration we plan to investigate here is illustrated in Figure \ref{fig:schematic}. An illuminating wavefront is transmitted through the 3D object and the plane P beyond the object is imaged by an afocal imaging system. We assume that the object field at the sensor plane is detected holographically by interference with a reference beam. Further, we will assume that the interference pattern(s) on demodulation is able to provide a complex-valued object field $u_{scat}(x,y)$ at the sensor using single-shot techniques \cite{singh2017single}. The field $u_{scat}(x,y)$ is treated as our data for the purpose of the reconstruction problem. For simplicity of treatment here, we assume the afocal imaging system to have a unit magnification.  
We discretize the 3D test object to have $N$ slices in the z-direction, which is also the nominal propagation direction of the axial illumination. Each of the z-slices is represented by a lumped transmission function:
\begin{equation}\label{eq:n3d}
    t_j (x',y') = a_j \exp[i \phi_j (x',y')], \;\;\;\; j = 1, 2, 3, ..., N.
\end{equation}
Here $a_j$ and $\phi_j$ represent the amplitude and phase of the $j$-th slice transmission function. The quantity $\phi_j$ contains the phase accumulated by the illumination wavefront due to the sample's refractive index over the interslice distance i.e., $\phi_j (x',y') = (2 \pi / \lambda) \int_{\Delta z} (n_j(x',y') - n_m) \ dz $, where $n_j$ and $n_m$ represent the refractive indices of the $j$-th slice and surrounding medium respectively and $\Delta z$ is interslice distance. The forward propagation of the illuminating wavefront through the 3D object is modeled via multi-slice beam propagation (MSBP) which involves sequential multiplication by the transmission functions $t_j (x',y')$ interspersed with free-space propagation between the adjacent slices. Effectively, we are lumping the medium spread over slabs of thickness $\Delta z$ into thin slices that are interspersed with free space. The MSBP model \cite{maiden2012ptychographic,chowdhury2019high,kamilov2016optical} is an advanced model that is also suitable for thick samples since it accounts for multiple scattering effects. We will denote the MSBP operation from the first illuminated slice of the object up to plane P by symbol $\hat{A}_l$: 
\begin{equation} \label{eq:sensingmat}
    u_{scat, l}(x,y) = \hat{A}_l \ n_{3D}(x,y,z).
\end{equation}
Here the index $l$ represents the $l$-th structured illumination and the variable $n_{3D}$ denotes the 3D object with slices having refractive index $n_j (x,y)$ as in (\ref{eq:n3d}). In the following discussion we will use a single or up to two different structured illuminations depending on the configuration being considered. For free-space propagation within the forward operation $\hat{A}_l$, we use the angular spectrum method denoted by operation $\hat{H}_z$ which represents free-space propagation of a field $u(x,y; z_0)$ by distance $z$. Denoting the 2D Fourier transform of the field $u(x,y; z_0)$ by $U(f_x,f_y; z_0)$, we have the relation:
\begin{equation}
u(x,y,z_0 + z) = \hat{H}_{z} u(x,y,z_0) = \mathcal{F}^{-1} \Big[ U(f_x,f_y; z_0) \, \exp(i\alpha z) \Big],
\end{equation}
where $\alpha = \sqrt{k^2 - 4\pi^2 (f_x^2 + f_y^2)}$ with $k = 2\pi/\lambda$ being the wave number associated with the illumination wavelength $\lambda$. A realistic afocal system as shown in Figure \ref{fig:schematic} will have its own impulse response due to the NA of the objective lens. In this work, we model this effect simply by band-limiting the field $u_{scat,l}(x,y)$ to a low-pass spatial frequency band decided by the NA. 
\subsection{Reconstruction methodology}\label{subsect:recon}
This section presents the reconstruction methodology for estimating slice-by-slice 3D refractive index from 2D scattered fields. 
The scattered field $u_{scat,l}(x,y)$ is a 2D representation of the solution $n_{3D}(x,y,z)$ that we are seeking. When the refractive index profile has multiple slices, the reconstruction of $N$ planes of the solution $n_{3D}(x,y,z)$ from $L$ scattered fields $u_{scat,l}(x,y)$ for $L < N$ is an incomplete data problem. Developments in the area of compressive sensing (CS) \cite{baraniuk2007compressive, candes2008introduction} over the last two decades allow us to solve such incomplete data problems by use of optimization algorithms that employ image sparsity as a constraint. For the purpose of setting up the optimization problem, we first define the data inconsistency objective function, which measures the least-square error between the estimated and measured 2D scattered field data for a given guess solution for $n_{3D}(x,y,z)$. This term is mathematically expressed as
\begin{equation}\label{eq:C1}
    C_1 = \sum_{l=1}^{L} C_{1l} = \sum_{l=1}^{L} \left \lVert u_{scat, l}(x,y) - \hat{A}_l \ n_{3D}(x,y,z) \right \rVert_{2}^{2}.
\end{equation}
Here $L$ is the total number of illuminations incident on the 3D sample and $\lVert.\rVert_2$ represents the $L_2$ norm. For enforcement of sparsity in the solution, we use the total variation (TV) penalty, which is a popular choice that is sufficient to illustrate the main points we wish to stress in this work. The TV penalty can be replaced by other penalties (e.g. Huber loss) if required. The TV penalty is defined as
\begin{equation}\label{eq:C2}
    TV(n_{3D}) = C_2 = \sum_{N_x} \sum_{N_y} \sum_{N_z} \sqrt{\lvert \nabla_{x} n_{3D} \rvert^2 + \lvert \nabla_{y} n_{3D} \rvert^2 + \lvert \nabla_{z} n_{3D} \rvert^2}.
\end{equation}
Here $N_x$, $N_y$, and $N_z$ represent the number of voxels of discretized 3D object in $x, y$, and $z$ directions, respectively. We use an alternating minimization type iterative scheme to reduce the two objective functions $C_1$ and $C_2$. The functional gradients $\nabla_{n_{3D}} C_{1l}$ and $\nabla_{n_{3D}} C_2$ required for the iterative updates are computed using the readily available automatic differentiation tools. We start the reconstruction by initializing $n_{3D} = n_{3D}^{(0)}$ as the guess solution. An adaptive gradient-descent scheme inspired by the Adaptive Steepest Descent - Projection onto Convex Sets (ASD-POCS) algorithm \cite{sidky2008image, malik2021computational, singh2017single} is then employed to reduce the two objective functions $C_1$ and $C_2$ alternatingly. For $(m+1)$-th iteration, we follow the steps described below.
\begin{enumerate}
\item {\bf Reduction of $C_1$}: The previous solution $n_{3D}^{(m)}$ is updated sequentially for all $L$ illuminations as:
\begin{equation}\label{eq:reduceC1}
    n^{(m,l+1)}_{3D} = n^{(m,l)}_{3D} - t_l \bigl[\nabla_{n_{3D}} C_{1l}\bigr]_{n_{3D} \ = \ n^{(m,l)}_{3D}},
\end{equation}
with $n^{(m,l=1)}_{3D} = n_{3D}^{(m)}$ and
$l$ ranging over $1, 2, ..., L$. The step size $t_l$ is determined using a backtracking line search \cite{armijo1966minimization}. The intermediate solution $n^{(m,int)}_{3D}$ is then equal to $n^{(m,l=L+1)}_{3D}$.
\item {\bf Compute distance $d_1$}: The change in the solution due to the error reduction step is estimated as:
\begin{equation}\label{eq:d1}
    d_1 = \lVert n^{(m,int)}_{3D} - n^{(m)}_{3D}\rVert_2.
\end{equation}
\item {\bf Reduction of $C_2$}: The TV-reduction step is then implemented recursively on the intermediate solution with a step size proportional to $d_1$ as:
\begin{equation}\label{eq:reduceC2}
    n^{(m)}_{3D_{k+1}} = n^{(m)}_{3D_{k}} - \beta d_1 \Biggl[ \frac{\nabla_{n_{3D}} C_{2}}{\lVert\nabla_{n_{3D}} C_{2}\rVert_2} \Biggr]_{n_{3D} \ = \ n^{(m)}_{3D_{k}}},
\end{equation}
with $k$ ranging over 0,1,2, ..., $K$ where $K$ represents fixed number of sub-iterations in each outer iteration and $n^{(m)}_{3D_{k=0}} = n^{(m,int)}_{3D}$. The parameter $\beta$ ranges in $(0,1)$ and for the present work, we used $\beta = 0.4$ and $K = 50$ in all the cases discussed.
\item {\bf Compute distance $d_2$}: The change in the solution due to TV reduction step can now be estimated as:
\begin{equation}\label{eq:d2}
    d_2 = \lVert n^{(m)}_{3D_{K-1}} - n^{(m,int)}_{3D}\rVert_2.
\end{equation}
\item {\bf Update $\beta$}: As per ASD-POCS, it is desirable to achieve $d_1 \approx d_2$, hence, if $d_1 < d_2$, we reduce the parameter $\beta$ by a fixed amount (for example, $\beta \rightarrow 0.95 \beta$) for the $(m+1)$-th outer iteration.
\item {\bf Update previous solution}: The solution $n^{(m+1)}_{3D}$ is now equal to $n^{(m)}_{3D_{K-1}}$. 
\end{enumerate}
The steps (i)-(vi) are then repeated until the change in the 3D refractive index solution is negligible. The relative reconstruction error, E($\%$), of the RI solution $n_{3D}$ with respect to ground truth RI $n^{(ground\ truth)}_{3D}$ can be estimated as:
\begin{equation}\label{RE}
\text{E(\%)} = \frac{\lVert n^{(ground\ truth)}_{3D} - n_{3D}\rVert_2}{\lVert n^{(ground\ truth)}_{3D}\rVert_2} \times 100 \%\ .
\end{equation}
The above reconstruction method for 100 outer iterations is uniformly applied to all the simulations in the present work. The reconstruction was performed using a Python code implemented on a computer with 12th Gen Intel(R) Core(TM) i7-1260P at 2.50 GHz 16 GB RAM CPU with 12 cores. The reconstruction time per iteration was $\sim 0.5$s. We have used the Google JAX library that offers efficient computations of steps required in implementing the iterative algorithm.

\subsection{Simulation study using axial Plane wave illumination}
\begin{figure}[tbp]
    \centering
    \includegraphics[width=0.5\textwidth]{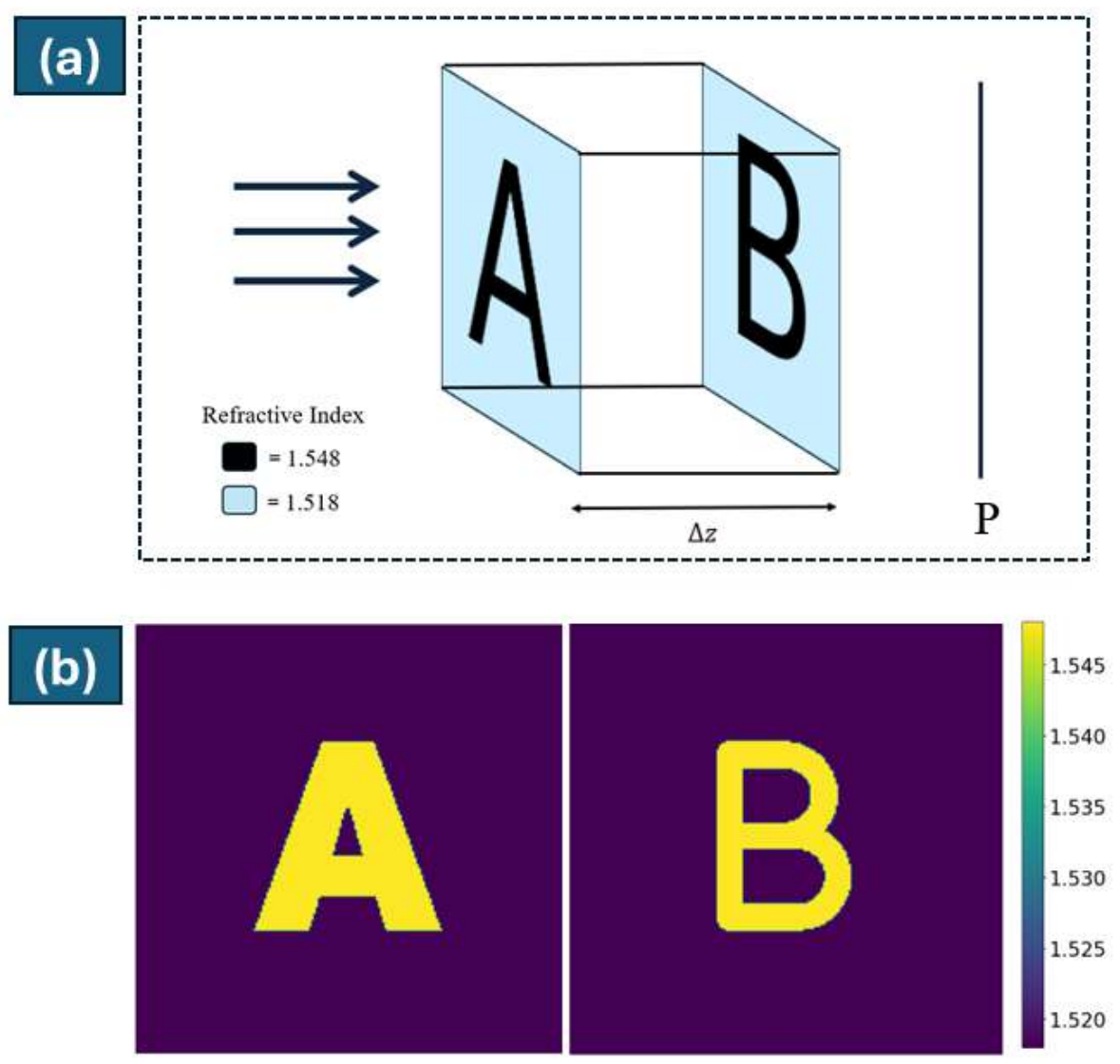}
    \caption{(a) Axial Plane wave illuminates 2-slice RI object with laterally overlapping letters A and B on individual planes separated by inter-slice distance $\Delta z$ and scattered field is detected at plane P and (b) Ground truth 2D RI profiles of individual slices.}
    \label{fig:Groundtruth}
\end{figure}

We first study the slice-by-slice recovery using plane wave illumination for a 2-plane RI object consisting of 2 letters A and B placed at two slices or z-planes as shown in Figure \ref{fig:Groundtruth}(a). We assume that the objects in the multiple slices have objects that are similar in structure. In general, RI is a complex quantity where real and imaginary components relate to the 3D phase and absorption from the sample, respectively. However, for the present work, we assume no absorption such that RI is a real quantity. We sample our object as a 3D computational box containing $200 \times 200 \times 2$ voxels. The physical size of the 3D box representing the object is equal to $200 \mu$m $\times \ 200 \mu$m laterally and the z-planes are separated by distance $\Delta z$ axially. The wavelength of the illumination field is assumed to be $650$nm. The letters A and B have RI equal to 1.548 immersed in a surrounding medium of 1.518 RI. The ground-truth 2D profiles for RI object are shown in Figure \ref{fig:Groundtruth}(b). The detection plane (plane P in reference to Figure \ref{fig:schematic}) is placed at a distance of $100 \mu$m from the object plane containing the letter B. The scattered field $u_{scat,l}(x,y)$ at plane P is treated as the data for the reconstruction. For practical consideration, we add independent Gaussian random noise realizations to the real and imaginary parts of the $u_{scat,l}(x,y)$ with standard deviation equal to 1/$\sqrt{2 N_0}$ where $N_0 = 5 \times 10^4$ is the average number of photons per pixel. The NA of the objective lens is assumed to be 0.3 for the present work. The object slices, as well as the detector, are assumed to sample the fields with 1$\mu$m lateral spacing.

\begin{figure}[tbp]
    \centering
    \includegraphics[width=0.85\textwidth]{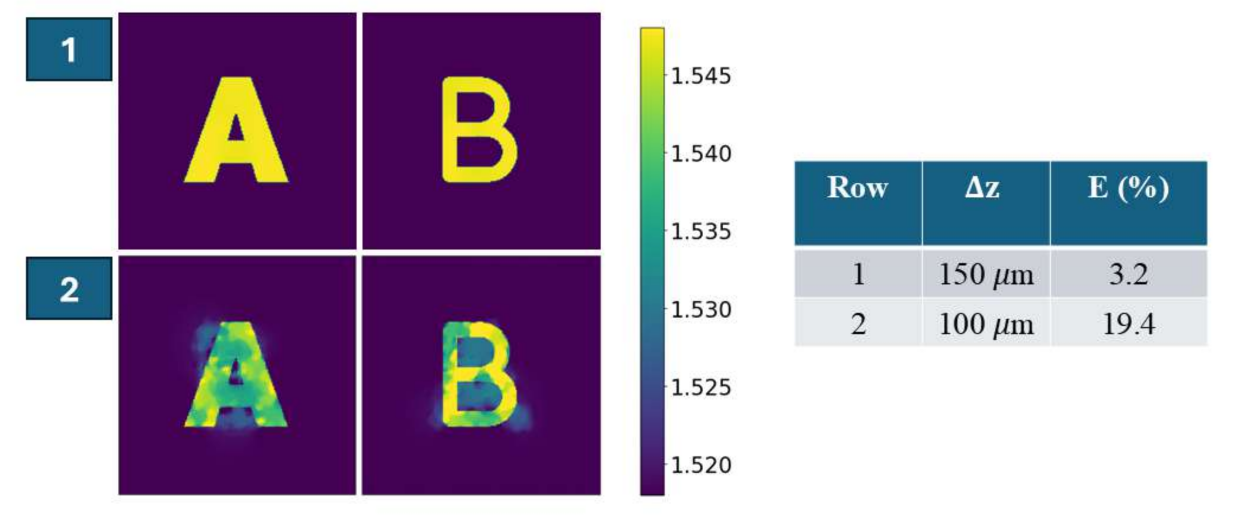}
    \caption{RI reconstructions of the two slices using single axial plane wave illumination for $\Delta z = 150 \mu$m and 100$\mu$m in rows 1 and 2, respectively. Row 2 shows cross-talk artifacts compared to Row 1. The relative reconstruction error is shown in the table on the right for both cases.}
    \label{fig:Planewavecase}
\end{figure}

As a first illustration, we use a single axial plane wave illumination and show the reconstruction of the two slices for two cases corresponding to $\Delta z$ =  $150 \mu$m and $100 \mu$m. Using the reconstruction method discussed in section \ref{subsect:recon}, we obtain the refractive index distribution for the two slices as shown in Figure \ref{fig:Planewavecase}. We observe that the reconstruction for the case $\Delta z = 150 \mu$m does not show cross-talk between the two slices (top row of Figure \ref{fig:Planewavecase}). The cross-talk is, however, present in the $\Delta z = 100 \mu$m case (bottom row of Figure \ref{fig:Planewavecase}). The reconstruction problem for a two-plane object with single illumination corresponds to an incomplete data problem with factor $N/L = 2$. For the type of objects used here, it is expected that this level of data incompleteness can be adequately handled by the TV sparsity penalty. Interestingly, the results in Figure \ref{fig:Planewavecase} illustrate that the quality of the reconstruction is not purely guided by sparsity considerations. As the iterative reconstruction progresses, we observe that for the $\Delta z = 100 \mu$m case, the features of A (or B) are retained in the B (or A) plane as we update the solution. As a result, the TV penalty on its own is unable to remove the cross-talk. On the other hand, the out-of-plane artifacts are much more diffused in the $\Delta z = 150 \mu$m case and are amenable to be mitigated by the TV penalty. Qualitatively, this suggests that in addition to object sparsity, the de-correlation of the fields propagated by distance $\Delta z$ is an important consideration for obtaining a cross-talk free reconstruction.

In earlier work \cite{rivenson2013reconstruction, stern2011conditions}, this point has been studied for Born approximation (or single scattering) based forward model. Our study, however, makes this point for the MSBP-based forward model. Under single scattering model \cite{rivenson2013reconstruction}, the quality of reconstruction is decided by the correlation between impulse responses for the multiple object slices in the detector plane. Under the MSBP model, it is not straightforward to talk about the impulse response for individual planes, as the response of a point object in a particular slice is dependent on the object distribution in other slices. It is, therefore, difficult to present our analysis in terms of the coherence parameter as used in \cite{rivenson2013reconstruction}. Instead, we base our discussion on the field de-correlation length that is dependent on the effective numerical aperture (or spatial bandwidth) of the fields. 

\subsection{Bandwidth ratio as a design guideline}\label{s33}
\begin{figure}[tbp]
    \centering
    \includegraphics[width=1.0\textwidth]{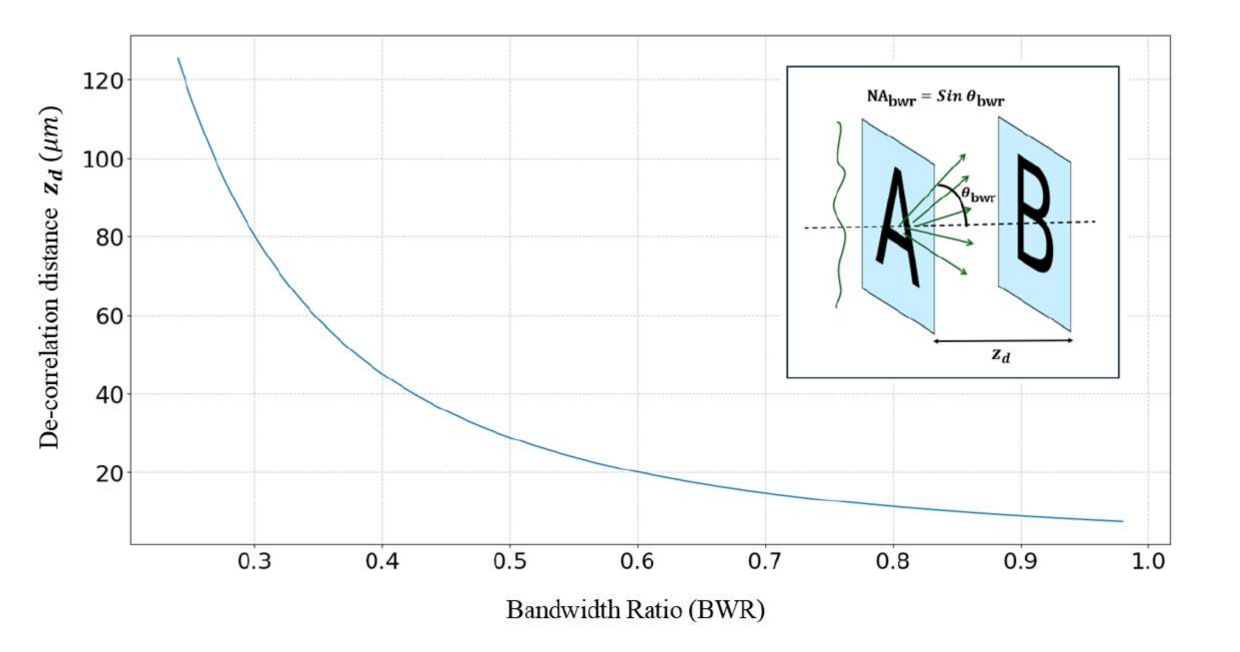}
    \caption{Design curve showing de-correlation distance $z_d$ as as function of BWR corresponding to exit field from slice A. The inset describes the relation of $\text{NA}_{bwr}$ with the spatial bandwidth of the exit field.}
    \label{fig:designguideline}
\end{figure}

In the MSBP model for the two-slice object, the B slice is illuminated by the propagated version (by distance $\Delta z$) of the exit wave from slice A. Our aim is to make sure that the exit field from slice A is adequately de-correlated with respect to the illuminating field at slice B. From the basic depth-of-focus considerations, it is clear that the required de-correlation distance is related to the effective spatial bandwidth of the exit wave from slice A rather than the NA of the downstream imaging system. For defining the effective bandwidth, we consider the energy concentration ratio for the exit field $u_{\text{exit}}(x,y)$:
\begin{equation}\label{eq:BWR}
     q = \frac{\int\limits_{\mathbf{f} \leq f_{\text{bwr}}} \ \ |U_{\text{exit}}(\mathbf{f})|^2\  d\mathbf{f}}{\int\limits_{\mathbf{f} \leq f_{\text{NA}}}\ \ |U_{\text{exit}}(\mathbf{f})|^2\  d\mathbf{f}} .  
\end{equation}
Here $U_{\text{exit}}(\mathbf{f})$ denotes the 2D Fourier transform of 
$u_{\text{exit}}(x,y)$, $\mathbf{f} = \sqrt{f^2_x + f^2_y}$, $f_{\text{NA}}$ = NA/$\lambda$ is the spatial frequency corresponding to system NA and $f_{\text{bwr}}$ represents the minimum spatial frequency for which $q$ exceeds 0.99. Further we introduce a quantity bandwidth ratio BWR = $f_{\text{bwr}} / f_{\text{NA}}$ for convenience in the following discussion. The exit field from slice A is expected to de-correlate adequately after a distance of $z_{d} \sim \lambda/(\text{NA}_{bwr})^2$, where $\text{NA}_{bwr} = \text{(NA)(BWR)}$. In Figure \ref{fig:designguideline} we plot this de-correlation distance as a function of BWR. The variable BWR values in our setup can be achieved by structured illumination of varying complexity.  
For the two-slice object reconstruction illustration in Figure \ref{fig:Planewavecase}, where we used plane wave illumination, the BWR corresponding to the exit wave of slice A is equal to 0.24 which corresponds to $z_d = 125 \mu$m. This explains the observation that the reconstruction for $\Delta z = 150\mu$m ($\Delta z > z_d$) has much better performance (in terms of error and cross-talk) compared to the $\Delta z = 100\mu$m ($\Delta z < z_d$) case. The plot in Figure \ref{fig:designguideline} can thus be treated as a design curve for achieving high-quality reconstruction. Our observation regarding $z_d$ also has parallels with the Fresnel similarity measure which was proposed in \cite{lohmann2005fresnel}. In the next section, we will illustrate this point further with axial structured illuminations.

\subsection{3D reconstruction using Axial randomly phase structured illumination(s)}
\begin{figure}[tbp]
    \centering
    \includegraphics[width=1.0\textwidth]{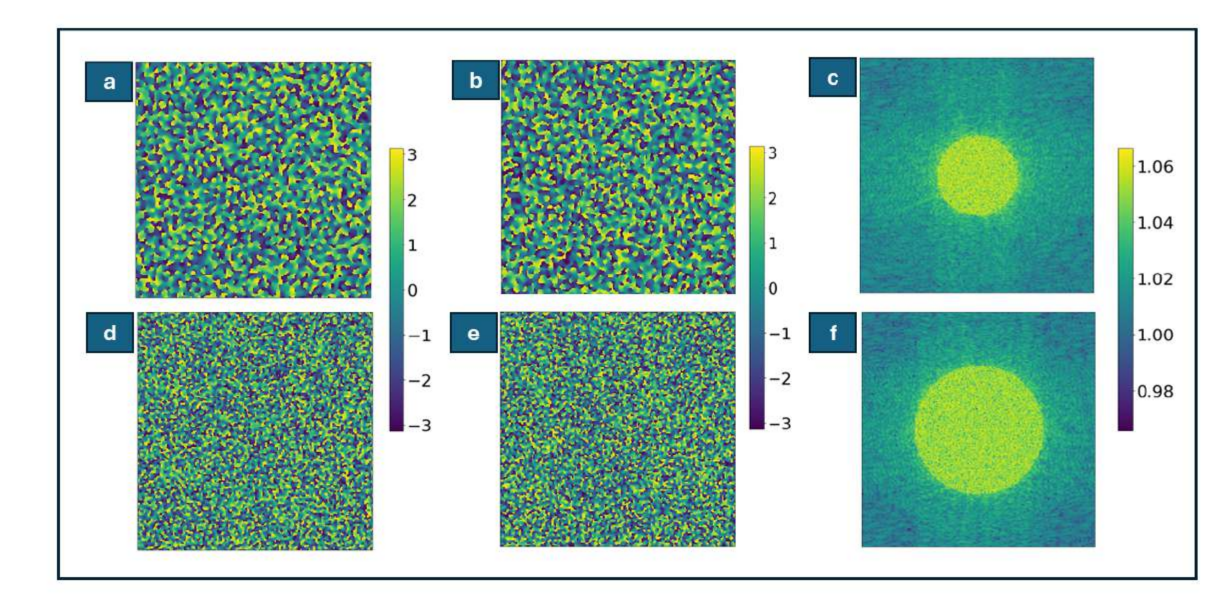}
    \caption{(a,d) Phase of structured illumination, (b,e) phase of scattered field and (c,f) its Fourier magnitude for BWR equal to 0.4 and 0.6 in top and bottom rows respectively. The interslice distance $\Delta z$ is equal to 20 $\mu$m for both cases.}
    \label{fig:fourierdomain}
\end{figure}

\begin{figure}[tbp]
    \centering
    \includegraphics[width=0.8\textwidth]{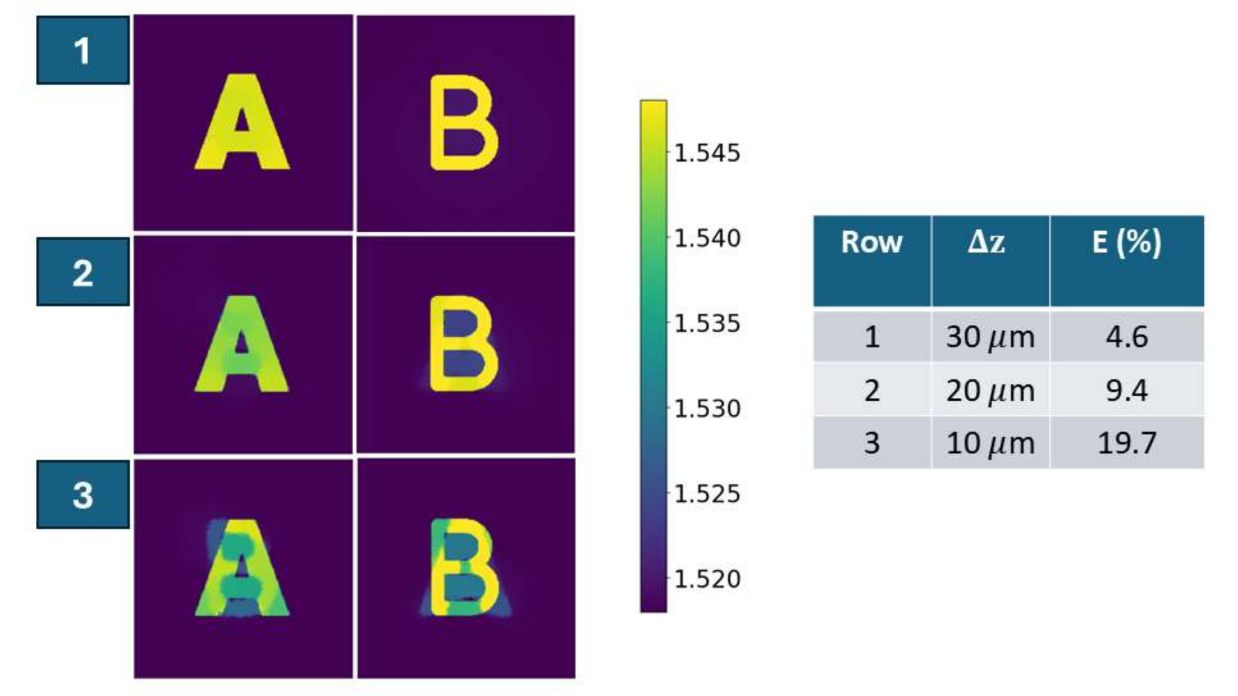}
    \caption{RI reconstructions of the two slices using single axial structured illumination with BWR = 0.6 for $\Delta z$ equal to $30 \mu$m, $20 \mu$m and $10 \mu$m in rows 1,2 and 3 respectively. The table on the right shows relative reconstruction error for each case.}
    \label{fig:neighbourhood}
\end{figure}

In this section, we use phase structured illumination instead of plane wave illumination. The BWR for the illumination is designed by low pass filtering of uniformly random phase patterns with phase distribution in $[0, 2\pi]$. The amplitude of the structured illumination is kept unity. We choose speckle-phase illumination since this type of illumination has nearly isotropic spectral coverage that results in uniform de-correlation for all spatial frequencies. This feature ensures that the reconstruction is not influenced by any preferred spatial frequency present in the object slices. The size of the speckle directly relates to the BWR ratio of the illumination. In Figure \ref{fig:fourierdomain}, we show the phase of structured illumination, the phase of scattered field, and its 2D Fourier magnitude for BWR = 0.4 (top row) and BWR = 0.6 (bottom row), respectively. It is evident that with the increase in BWR, the illumination transitions from near plane wave (low BWR) to a highly structured illumination, and the space-bandwidth product of the measured data increases accordingly. It is interesting to note that, random illuminations have previously been used to eliminate cross-talk in the 3D holographic displays \cite{makey2019breaking}. Further, such random structure is typically considered to be a good choice for encoding information in CS literature through coded-aperture \cite{horisaki2014single} or coded illumination for 2D and 3D phase microscopic imaging \cite{yeh2019speckle, horisaki2016single, horisaki2019diffusion}.
Following the design curve in Figure \ref{fig:designguideline}, we choose a structured illumination with BWR equal to 0.6 that suggests a de-correlation distance of $z_d$ equal to 20 $\mu$m. In Figure \ref{fig:neighbourhood}, we show the slice-by-slice recovery for three inter-slice distances equal to 30, 20 and 10 $\mu$m respectively. We observe that at $\Delta z = 30 \mu$m (row 1), cross-talk free reconstruction is achieved with reduced relative error. If we use $\Delta z = 20 \mu$m (row 2), we are exactly on the curve for BWR = 0.6 and we see a faint cross-talk between the slices. For $\Delta z = 10 \mu$m (row 3), the cross-talk error is increased. This observation once again validates the usefulness of the design curve (Figure  \ref{fig:designguideline}) in deciding an optimal inter-slice distance for a given BWR of the exit wave and vice-versa. 

\begin{figure}[tbp]
    \centering
    \includegraphics[width=1.0\textwidth]{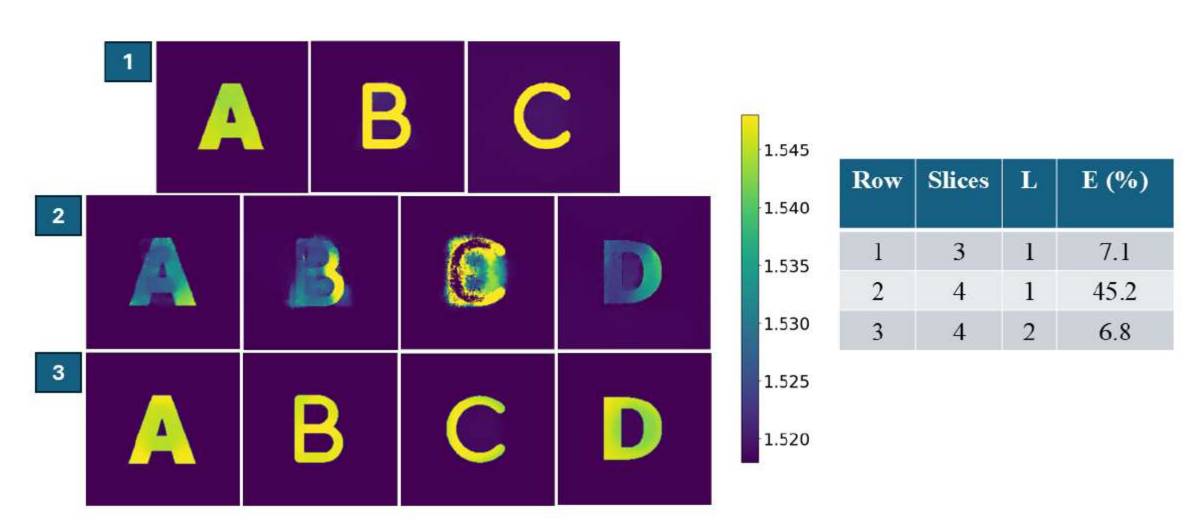}
    \caption{RI reconstructions for $\Delta z = 30 \mu$m using axial structured illumination(s) with BWR = 0.6. Row 1 corresponds to 3-slice object reconstruction with one illumination. Rows 2 and 3 show 4-slice object reconstruction using one and two illumination(s), respectively. The table on the right shows relative reconstruction error for each case where L represents the number of illuminations used.}
    \label{fig:CS}
\end{figure}

Having selected the inter-slice distance $\Delta z$ to be greater than the de-correlation distance $z_d$, we now study the feasibility of good quality tomographic reconstruction for the same configuration in row 1 of Figure \ref{fig:neighbourhood} i.e., $\Delta z = 30 \mu$m, but now with increased number of slices in the 3D object. It is important to mention that with a sufficient BWR of the exit wave from the first slice, it is ensured that the de-correlation condition will be satisfied for the subsequent slices as well. Figure \ref{fig:CS}, row 1 shows recovery for 3 slices A, B and C, where we observe a reconstruction with minimal cross-talk. However, when 4 slices (row 2) are introduced (i.e., A, B, C, and D), the reconstructed slices show noticeable cross-talk artifacts even though the same $\Delta z$ is used. The worsening of solution quality here may be attributed to the dimensionality mismatch between the number of unknowns in the 3D volume to be reconstructed and the number of pixels in the measured 2D data frame for a single structured illumination. The sparsity constraint (in the form of a TV penalty) is not sufficient for handling the dimensionality mismatch. By using one additional illumination with the same complexity (BWR = 0.6), we see a high-quality reconstruction again in row 3 with reduced artifacts and relative data domain error. The illustrations in Figures \ref{fig:neighbourhood} and \ref{fig:CS} clearly show the role of structured illumination and sparsity-penalty in achieving cross-talk free reconstruction. For a given tomographic configuration and nature of 3D object, a combination of both the design curve in Figure \ref{fig:designguideline} and sparsity penalty is thus important in deciding the minimal number of axial structured illuminations required. 

\section{Discussion and Conclusion}\label{s4}

The present study aims to understand the feasibility of implementing a phase tomography system by employing axial structured illumination(s). The current phase tomography systems are mostly designed to have multi-angle plane wave illuminations to provide data diversity. The multi-angle illumination model aligns very well with the early ideas on ODT. The reconstruction algorithms in the current phase tomography literature are, however, moving away from direct reconstruction (based on the Fourier diffraction theorem) towards iterative reconstruction schemes. Iterative methods naturally allow better modeling of the forward operator and allow the use of image domain penalty functions and, therefore, have validity beyond the Born approximation. The main thought behind our work is that if iterative reconstruction is to be ultimately used, the system geometry can have more flexibility as long as the measurement scheme provides diversity offered by multi-angle illuminations. From a hardware simplification perspective, the replacement of multi-angled illuminations by axial structured illumination (consisting of a combination of plane waves) offers attractive possibilities. 

In order to design an ASIT system, the role of illumination diversity and sparsity constraints used by iterative reconstruction schemes need to be identified clearly, as we have presented here. Firstly, by using a single illumination (plane wave or structured), we show that the quality of sparsity-based reconstruction is strongly dependent on the inter-slice separation in the 3D volume of interest, even for the case of a 2-slice object. In order to understand this observation further, we utilized the concept of de-correlation distance $z_d$ and presented a design curve that allows us to select the appropriate inter-slice distance for a given spatial bandwidth (or BWR) of the exit wave from the first slice in the object. The use of BWR as a guideline for selecting the minimum required $\Delta z$ makes the conclusions of our study nearly independent of the nature of the object being imaged. As per the design curve (Figure \ref{fig:designguideline}), BWR = 1 case allows us to make $\Delta z$ comparable to the depth-of-focus of the imaging system being used. 
Having selected this inter-slice distance, we then explore the quality of reconstruction as the number of slices in the object are progressively increased. This later study shows the allowable data mismatch that the sparsity penalty can handle effectively and when additional illuminations may have to be employed for achieving cross-talk free reconstruction. Overall, the simulation study provides a systematic methodology for designing an axial structured illumination phase tomography system as an alternative to multi-angle illumination-based systems. From the system implementation perspective, the structured illumination(s) used need to be calibrated for use in the iterative reconstruction. But once calibrated, such a system can potentially be leaner than a multi-angled system.

In the present work, we used MSBP as the forward model for wave propagation inside the sample, which is appropriate when the RI contrast is small. However, if there is a significant RI contrast, advanced models such as the Multi-layer Born model \cite{chen2020multi} (accounts for back-scattering within the sample) may have to be used, and the design analysis presented here may have to be repeated for this more involved forward modeling. In the MSBP (or the multi-layer Born) models, the "measurement matrix" cannot be de-coupled from the object as is the standard practice in theoretical developments associated with the CS theory. The exact CS results on the reduction in a number of measurements (as compared to a number of unknown voxels in the 3D volume) cannot, therefore, be employed in our analysis directly but can only serve as a nominal guideline. The CS-based analysis of phase tomography problem in terms of coherence parameter $\mu$ as performed in \cite{rivenson2013reconstruction} for the Born model is difficult to apply when MSBP is used. However, taking a cue from this early study we base our discussion on the BWR and the corresponding de-correlation distance $z_d$ as presented in Section \ref{s33}. Without getting into any tedious evaluations associated with coherence parameter, this methodology presents a practical approach to designing the nature and number of structured illuminations. Another degree of simplicity in our analysis is the assumption that all slices of the tomographic object have similar complexity (or sparsity). This assumption is not too restrictive and if it is not valid for a particular class of 3D objects, the methodology presented here can be potentially modified appropriately. Our future work will explore the use of designer axial illuminations that are minimally correlated with each other to further improve the data diversity. Experiments based on the present work will be reported as well.

\section*{Acknowledgements}
NG: Prime Minister Research Fellowship; KK: Abdul Kalam Technology Innovation National Fellowship (INAE).

\bibliographystyle{unsrt}  


\end{document}